\documentclass[aps,prl,reprint]{revtex4-1}
\usepackage{amsmath,amssymb,amsthm,amsfonts}
\usepackage{color}
\usepackage{hyperref}
\usepackage{graphicx}
\usepackage[utf8]{inputenc}

\begin{document}
\title{Determination of CFT central charge by the Wang-Landau algorithm}
\author{P.A. Belov}
\author{A.A. Nazarov}
\affiliation{Department of Physics, St. Petersburg State University,\\ Ulyanovskaya 1, 198504 St.~Petersburg, Russia}
\author{A.O. Sorokin}
\affiliation{Petersburg Nuclear Physics Institute NRC Kurchatov Institute,\\ Orlova roscha, 188300 Gatchina, Russia}

\begin{abstract}
  We propose a simple method to estimate the central charge of the conformal field theory
  corresponding to a critical point of a two-dimensional lattice model from Monte Carlo simulations. The main idea is to
  use the Wang-Landau flat-histogram algorithm, which allows us to obtain the free energy of a
  lattice model on a torus as a function of torus radii. The central charge is calculated with a
  good precision from a free energy scaling at the critical point. We apply the method to
  the Ising, tricritical Ising (Blume-Capel), Potts and site-diluted Ising models, and also discuss
  estimation of conformal weights.
\end{abstract}

\maketitle

\section{Introduction}
\label{sec:introduction}
A key aspect of the modern theory of critical phenomena is the similarity hypothesis formulated a
half century ago (see \cite{kadanoff1967static} for a historical review). The similarity is
manifested as a "scaling law", determining exponents of singular behavior for some important
thermodynamic quantities such as the specific heat and susceptibility. Kadanoff's generalization of
the similarity hypothesis \cite{kadanoff1966introduction} inspired Wilson to apply the
renormalization group approach \cite{wilson1971renormalization}. These ideas made a breakthrough in
the theory of critical phenomena \cite{fisher1998renormalization,stanley1999scaling}.

Another efficient approach is based on the conformal field theory (CFT). The scale invariance
together with the homogeneity of the ground state indicates an existence of an additional symmetry
at the critical point --- the conformal invariance \cite{Polyakov:1970xd}. In two dimensions, this
symmetry is especially powerful since it manifests the infinite-dimensional Virasoro algebra
\cite{BPZ}. The two-dimensional CFT provides a wealth of information about critical behavior, in
particular, it makes possible to compute analytically or numerically a lot of observables such as
multi-point correlation functions \cite{difrancesco1997cft}.

Conformal field theories are classified by the central charge $c$, that characterizes
representations of the Virasoro algebra
\begin{equation*}
  \label{eq:2}
\left[L_n,L_m\right]=(n-m)L_{n+m}+\frac{c}{12}(n^3-n)\delta_{n+m,0},
\end{equation*}
where $L_{n,m}$ are generators of the algebra, $\delta_{n+m,0}$ is the Kronecker delta. Generally,
CFT models have a small number of parameters, namely the central charge and conformal weights of the
primary fields, $h_{i}$, which are the eigenvalues of the operator $L_{0}$. Conformal weights
directly determine anomalous dimensions and other critical exponents, defining the universality class
of the critical point.

In contrast to higher dimensions, in two dimensions critical exponents are less significant, and the
central charge is the most important quantity defining the universality class. This can be seen by
the example of the Ashkin-Teller model \cite{ashkin1943statistics}, where critical exponents depend continuously on the
parameter of the model within the same symmetry class and the same value of the central charge $c=1$
\cite{kadanoff1979correlation}. Such a behavior is typical for models with
$c\geq 1$ \cite{zamolodchikov1986two,zamolodchikov1987conformal}.

Generally speaking, it is not easy to establish  a correspondence between CFT with a particular value of the
central charge and a symmetry class. On the one hand,
internal symmetries are not fully explored even for so-called minimal models (with a finite number of
Virasoro algebra modules). On the other hand, it is {\it a priori} unknown what CFT describes a
critical behavior of a given lattice model, even if a symmetry class of the model is known. Such a
situation is well represented, e.g., in frustrated spin systems.  Critical behavior of such systems is
often a subject of debate and controversy, for example, the controversy in $\mathbb{Z}_2\otimes
SO(2)$ symmetry class discussed for three decades (see \cite{korshunov2006phase,sorokin2012chiral} for a review). 

Another non-trivial example considered in this paper is the site-diluted Ising model. Using the symmetry
arguments, one expects that the critical behavior is the same as in the pure Ising model with
$c=1/2$. But previous investigations have given contradictory results
\cite{dotsenko1983critical,shalaev1984correlation,andreichenko1990monte,kim1994critical,najafi2016monte}.

In the non-trivial cases, a determination of the central charge is a complicated and important
problem. Several approaches have been proposed earlier by different authors
\cite{lauwers1991estimation,bastiaansen1998monte,feiguin2007interacting}. The first
of these methods \cite{lauwers1991estimation} is based upon the deformation of the CFT minimal model
by the magnetic field, so there are a lot of cases where it is not applicable. The second approach
\cite{bastiaansen1998monte} requires careful simulations and precise multi-parametric numerical
fitting of the conformal weights. The third method is only applicable to the one-dimensional
integrable models \cite{feiguin2007interacting}. 

In this paper, we propose a more straightforward approach which does not have mentioned
shortcomings. The main idea is to use the Wang-Landau flat-histogram algorithm
\cite{wang2001efficient}, which allows to obtain the free energy of a lattice model on a torus
as a function of torus radii. The central charge is calculated with a good precision from a free
energy scaling at the critical point. To test our method, we apply it to several models with the
well-known values of the central charge. We consider the Ising, tricritical Ising, three- and
four-state Potts models. As a non-trivial case, we report our results on the site-diluted Ising
model, that are presented in detail in a separate paper \cite{Belov:2016lak}.

\section{Theory and method}
\label{sec:theory}

\subsection{Wang-Landau algorithm}
\label{sec:Wang-Landau algorithm}

We consider lattice models formulated on regular lattices with lattice variables (usually spins) at
the lattice vertices. The Wang-Landau algorithm \cite{wang2001efficient,landau2014guide} simulates
the energy distribution $\rho(E)=e^{g(E)}$ of the lattice model. The energy range is split in some
number of intervals (which can coincide with the number of discrete energies). The algorithm starts
with a random lattice configuration, an empty array of logarithms of energy densities
$g(E_{1}),\dots,g(E_{n})$, an empty visitation histogram $h(E_{1}),\dots,h(E_{n})$ and some initial
value (usually 1) of constant $a$. Then, one lattice site and a new value of the lattice variable at
this site are chosen randomly. New state with the changed energy value is accepted with the
probability $e^{g(E_{new})-g(E_{old})}$, at the same time the visitation number $h(E)$ is increased
by 1 and $g(E)$ is increased by $a$. This procedure is repeated until the visitation histogram is
relatively flat. Then the value of $a$ is divided by 2, the histogram is emptied and next step of
the algorithm begins. Usually about 25-30 such steps are done in simulations. Having the energy
distribution $\rho(E)$, the partition function is given by
\begin{equation}
  \label{eq:8}
  Z=\sum_{E_{i}} \rho(E_{i}) e^{-\frac{E_{i}}{T}}.
\end{equation}

\begin{figure}[t]
  \centering
  \includegraphics[width=0.94\linewidth]{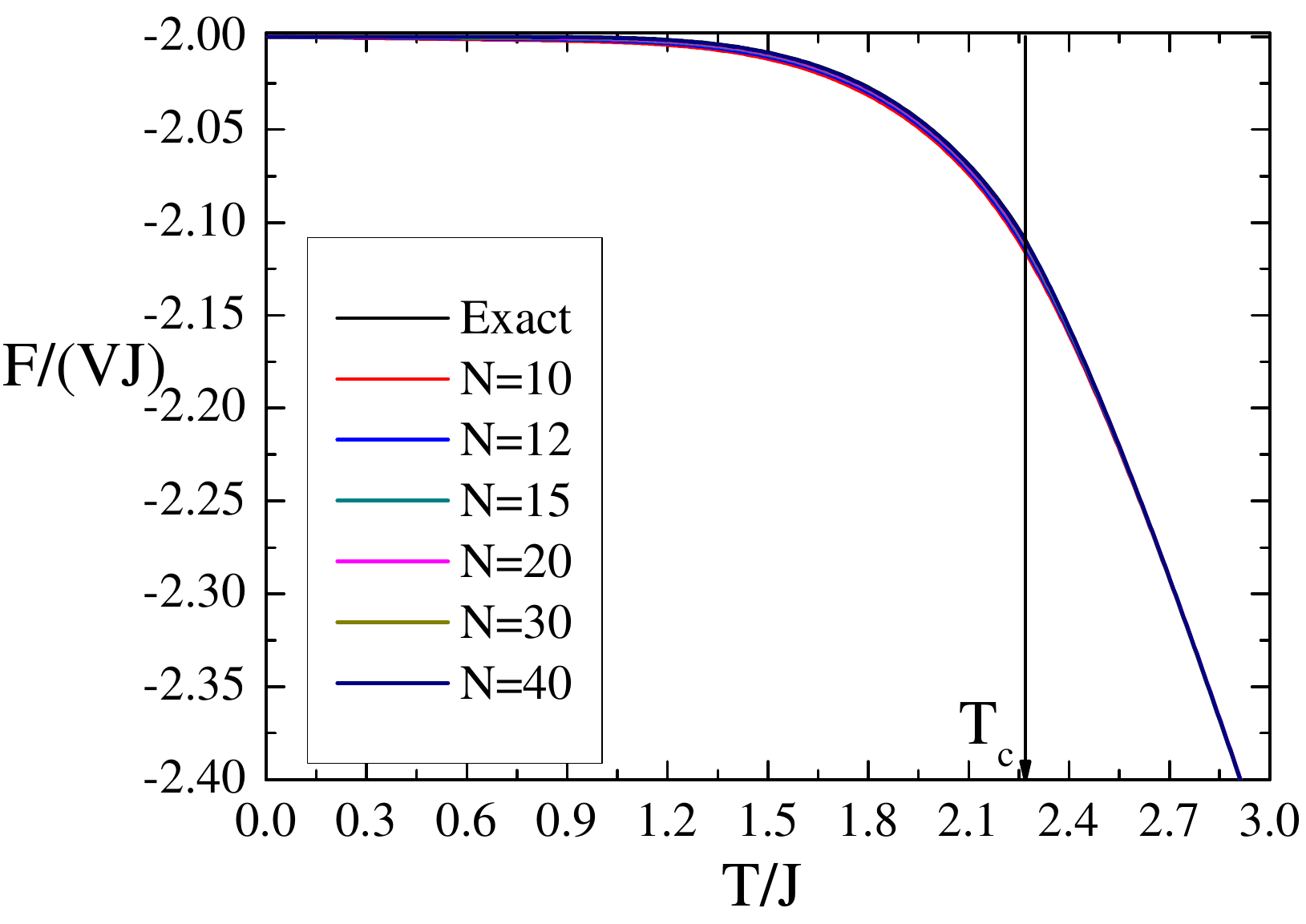}
  \caption{\label{FTIsing} Free energy density in the Ising model.}
\end{figure}
Availability of the partition function in simulations is the crucial difference between the
Wang-Landau and other Monte-Carlo methods, such as Metropolis
\cite{metropolis1953equation,landau2014guide} or Wolff algorithms \cite{wolff1989collective}. Having
the partition function, we can consider the free energy density
\begin{equation}
  \label{eq:10}
  f(T)=-\frac{T}{V}\log Z(T),
\end{equation}
where $V$ is a volume of a system. Typical thermal dependence of the free energy density is shown in
Fig. \ref{FTIsing} by the example of the Ising model on a square lattice. For comparison with the
Monte Carlo results, we also show in Fig. \ref{FTIsing} the exact result, obtained from the
Onsager's solution \cite{onsager1944crystal}.

\subsection{Central charge}
\label{sec:sentral charge}

In the conformal field theory there is the well-known relation connecting the free energy $f(N)$ on
an infinite cylinder of circumference $N$ with the free energy $f_0$ on a plane (see, for example
\cite{difrancesco1997cft}, Chapter 5):
\begin{equation}
  \label{eq:11}
  f(N)=f_0-\frac{\pi c}{6 N^{2}}.
\end{equation}
Since the central charge appears in equation \eqref{eq:11}, it can be used to extract the central
charge value from the simulation data. In order to do this, we simulate a model on a torus of
circumferences $N$ and $M$ to obtain the free energy density $f(N,M)$. Then we need to extrapolate the free
energy value as $M\to\infty$ to calculate the free energy on a cylinder $f(N)=\lim_{M\to\infty} f(N,M)$, which is used to obtain the central charge by fitting the equation~\eqref{eq:11}. 
To do this extrapolation, we need to know the detailed behavior of $f(N,M)$ on $M$.

Let us consider the CFT partition function on a torus of circumferences $M$ and $N$. In contrast to Bastiaansen and Knops
\cite{bastiaansen1998monte} who had to study a behavior on a  ``skew'' torus, we consider straight rectangular torus. The modular
parameter of this torus is given by the ratio of two periods:
\begin{equation}
  \label{eq:2}
  \tau=i\frac{M}{N}.
\end{equation}
In the usual CFT quantization, one needs to choose time direction. We take it to be along the
period $M$ of a torus. The Hamiltonian is the generator of time translations and is given by a sum
of Virasoro generators $L_{0}, \bar L_{0}$ \cite{difrancesco1997cft}:
\begin{equation}
  \label{eq:3}
  H=\frac{2\pi}{N} (L_{0}+\bar L_{0})-\frac{\pi c}{6 N},
\end{equation}
the additional term with the central charge $c$ appears from the conformal mapping from a plane. 

We can consider the exponent of the Hamiltonian as a row-to-row transfer matrix. Translating
from row to row along the time direction $M$ times we get the partition function on the torus
\begin{equation}
  \label{eq:1}
  Z=\sum_{j} \left<j\right|e^{-M H} \left|j\right>.
\end{equation}
The sum here runs over states in the Hilbert space, which is in turn a direct sum of Virasoro algebra
modules $V_{i}$ generated by the primary fields $\varphi_{i}$ and parametrised by the conformal weights
$h_{i}, \bar h_{i}$. A conformal field theory always contains the identity operator with the
conformal weight $h_{0}=0$. The conformal weights of the other primary fields are greater then zero.
We consider the minimal non-trivial conformal weight $h_{min}$ that usually lies between zero and one: $0<
h_{min},\bar h_{\bar min}< 1$. We can choose the basis of Virasoro eigenstates
$\left|h_{i}+m_{i},\bar h_{i}+\bar m_{i}\right>$ in the Hilbert space, where $m_{i},\bar m_{i}$ are
positive integers:
\begin{equation}
  \label{eq:4}
  \begin{array}{l}
    L_{0}\left|h_{i}+m_{i},\bar h_{i}+\bar m_{i}\right>=(h_{i}+m_{i})\left|h_{i}+m_{i},\bar h_{i}+\bar
    m_{i}\right>\\
  \bar L_{0}\left|h_{i}+m_{i},\bar h_{i}+\bar m_{i}\right>=(\bar h_{i}+\bar m_{i})\left|h_{i}+m_{i},\bar h_{i}+\bar
    m_{i}\right>    
  \end{array}
\end{equation}

It is customary to use the parameter $q=\exp(2\pi i \tau)$. In our case of the straight rectangular
torus the parameter $q$ is real, $q=\bar q=\exp\left(-2\pi M/N \right)$. Then for the partition function
we obtain
\begin{multline}
  \label{eq:5}
  \frac{Z(q)}{Z_{0}}=q^{-\frac{c}{24}} \bar q^{-\frac{c}{24}}\sum_{j} n_{j} q^{h_{j}+m_{j}}\bar q^{\bar h_{j}+\bar
    m_{j}}=\\=\sum_{i,\bar i} {\cal M}_{i,\bar i}\chi_{i}(q) \bar\chi_{\bar i}(\bar q),
\end{multline}
where $j$ runs over all states,, $n_{j}$ the multiplicity of the secondary state, $i, \bar i$ run
over the primary states, $\chi_{i}(q)=q^{h_{i}-\frac{c}{24}}\sum_{n\geq 0}d_{i}(n)q^{n}$ is the
character of the Virasoro algebra module and ${\cal M}_{i,\bar i}$ is the multiplicity of the
representation $V_{i}\otimes\bar V_{\bar i}$. The partition function is defined up to normalization
$Z_{0}$ that is interpreted as a partition function on a plane. The multiplicities
$\mathcal{M}_{i,\bar i}$ are non-negative integers that are constrained by the modular invariance of
the partition function. It is important to note that $\mathcal{M}_{0,0}=1$ since, as we have already
mentioned, the CFT always contains the identity field with $h_{0}=\bar h_{0}=0$.

The partition function in the conformal field theory does not depend explicitly on temperature since
CFT is applicable only in thermodynamic limit at critical point. Substituting equation \eqref{eq:5}
to the equation \eqref{eq:10} at the critical point we obtain
\begin{multline}
  \label{eq:13}
  f(N,M)=f_{0} - \frac{T_{c}\pi c}{6
    N^{2}}-\frac{T_{c}}{MN}\log\left[1+\phantom{\sum_{i\neq 0,\bar i\neq 0}}\right.\\ \left.\sum_{i\neq 0,\bar i\neq 0} \mathcal{M}_{i,\bar i} q^{h_{i}+\bar h_{\bar
        i}} \left(\sum_{n\geq 0} d_{i}(n) q^{n}\right)\left(\sum_{m\geq 0} d_{\bar i}(m)
      q^{m}\right)\right.\\ +\left.\sum_{n\geq 1}d_{0,0}(n)q^{n}\right].
\end{multline}
The unit in the logarithm appears due to the identity field, and we have moved the contributions
from the secondary states of the identity field $\sum_{n\geq 1}d_{0,0}(n)q^{n}$ to the right. Note
that if $M\geq N$ the parameter $q=e^{-\frac{2\pi M}{N}}$ is small, so we can expand the logarithm
holding only leading contributions with $q^{h_{i}}$, for $i$, such that $0<h_{i}<1$:
\begin{multline}
  \label{eq:mainfit}
  f(N,M)=f_{0} - \frac{T_{c}\pi c}{6
    N^{2}}-\frac{T_{c}}{MN}\left[\sum_{i\neq 0,\bar i\neq 0} \mathcal{M}_{i,\bar i} q^{h_{i}+\bar h_{\bar i}} \right].
\end{multline}
This formula is the main one for our estimation of the central charge and the conformal weights.

\subsection{Inaccuracies}
\label{sec:inaccuracies}

\begin{figure}[t]
  \centering
  \includegraphics[width=0.95\linewidth]{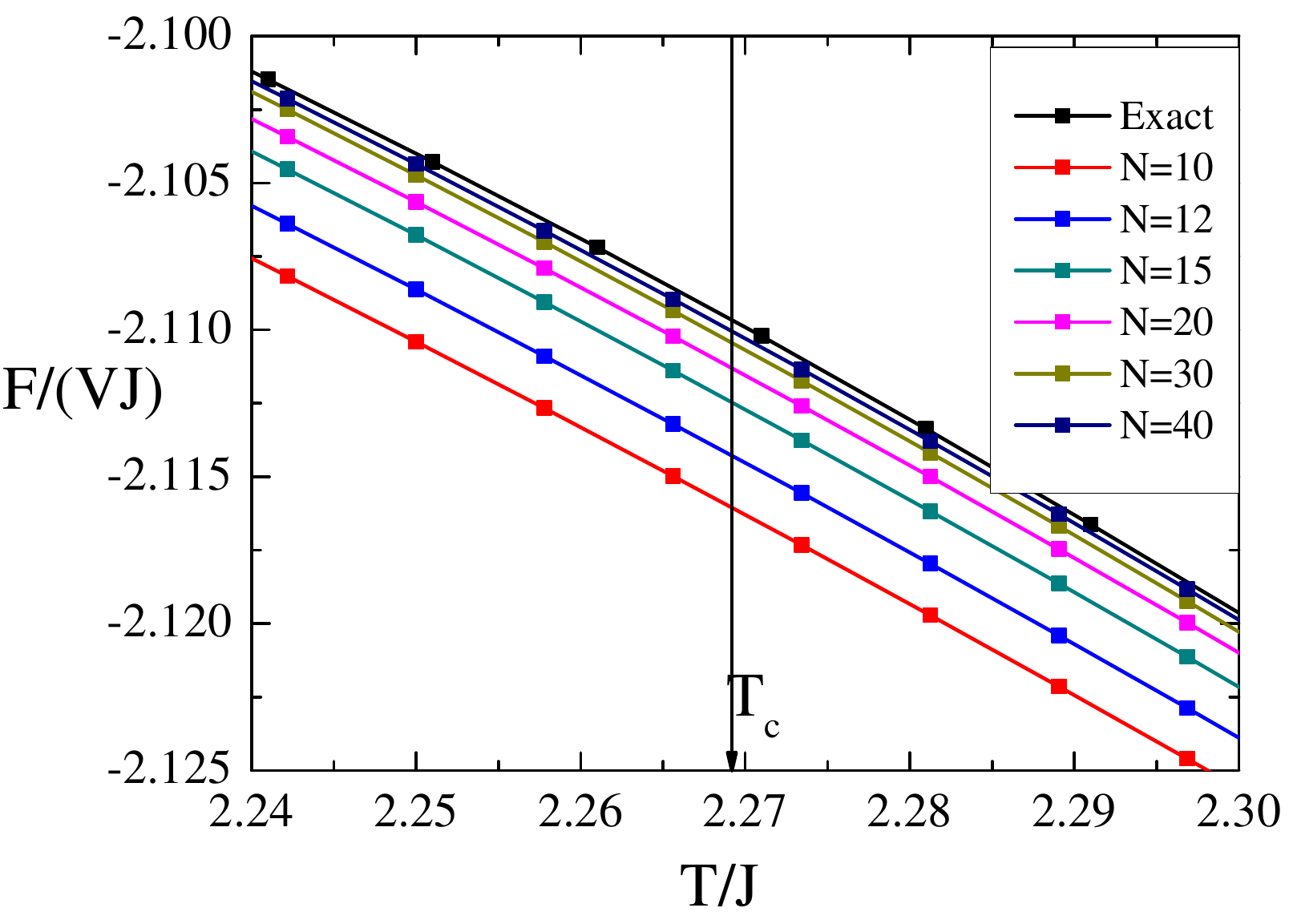}
  \caption{\label{FTcIsing} A thermal dependence of the free energy for different values of $N$  near the critical point.}
\end{figure}
There are several sources of inaccuracies. The first of them has the origin from the algorithm
dependence on the conditions of the visitation histogram flatness and number of iterations (steps).
Using 30 steps and 20\% difference between a visitation number of each energy states and the average
one, we estimate the free energy density with no more than $0.5\%$ inaccuracy
\cite{wang2001efficient,landau2014guide}.

The second one follows from the estimation error of the critical temperature. But as one can see in
Fig.~\ref{FTcIsing}, finite-size scaling corrections to the free energy depend on temperature
very weakly and remain actually the same in a rather wide range of temperatures near the critical
point. In practice, the critical temperature can be estimated quite precisely, so the second source
of inaccuracies is insignificant compared to the rest sources.

The procedure of data fit by formula \eqref{eq:mainfit} may be performed in several ways depending on the accuracy of data:
\begin{enumerate}
\item The multiple fit procedure with the estimation of a value $f(N)$ and a few (two or more)
  exponents $q^{\Delta_{i,\bar i}}$, with $\Delta_{i,\bar i}=h_i+\bar{h}_{\bar i}$. Note that a
  multiplicity $\mathcal{M}_{i,\bar i}$ is small non-negative integer.
\item The multiple fit procedure with estimation of a value $f(N)$ and single exponent
  $q^{\Delta_{min}}$, where $\Delta_{min}$ is the minimal anomalous dimension. Such a procedure becomes
  correct if $M>N$ when other exponents $q^\Delta$ become negligible. We use this variant of the fit
  procedure in the current study.
\item The simple fit procedure with estimation of a value $f(N)$ using an arbitrary exponent
  $q^{\Delta_r}$, with $\Delta_r\in[0,1]$. Such a procedure is only accurate if
  $M\gg N$ (say, e.g., $M\approx10N$), where the dependence on $\Delta_{r}$  is very weak.
\item The simplest and fastest procedure of the central charge estimation is a guess that
  $f(N)\approx f(N,AN)$, where $A$ is a large integer ($A\geq10$). This procedure is valid if $M\gg
  N$ due to the exponential smallness of corrections.
\end{enumerate}

In the table placed below, we show the estimations of the free energy value $f(N)$ with $N=10$.
The exponent $\Delta$ and the central charge obtained by the last three methods for the Ising model. The uncertainty of the last decimal digit is given in brackets.
\begin{equation}
  \label{eq:16}
  \begin{array}{c|c|c|c}
    \hline
    \hline
     &  \mbox{method 2} & \mbox{method 3}& \mbox{method 4}\\
    \hline
    \Delta & 0.14(3) & 1 & \mbox{not used} \\
    f(10) & -2.1158(6) & -2.115(1) & -2.1157(6) \\
    c &  0.50(2) & 0.46(6) & 0.51(3) \\
    \hline
    \hline
  \end{array}
\end{equation}
\begin{figure}[b]
  \centering
  \includegraphics[width=0.95\linewidth]{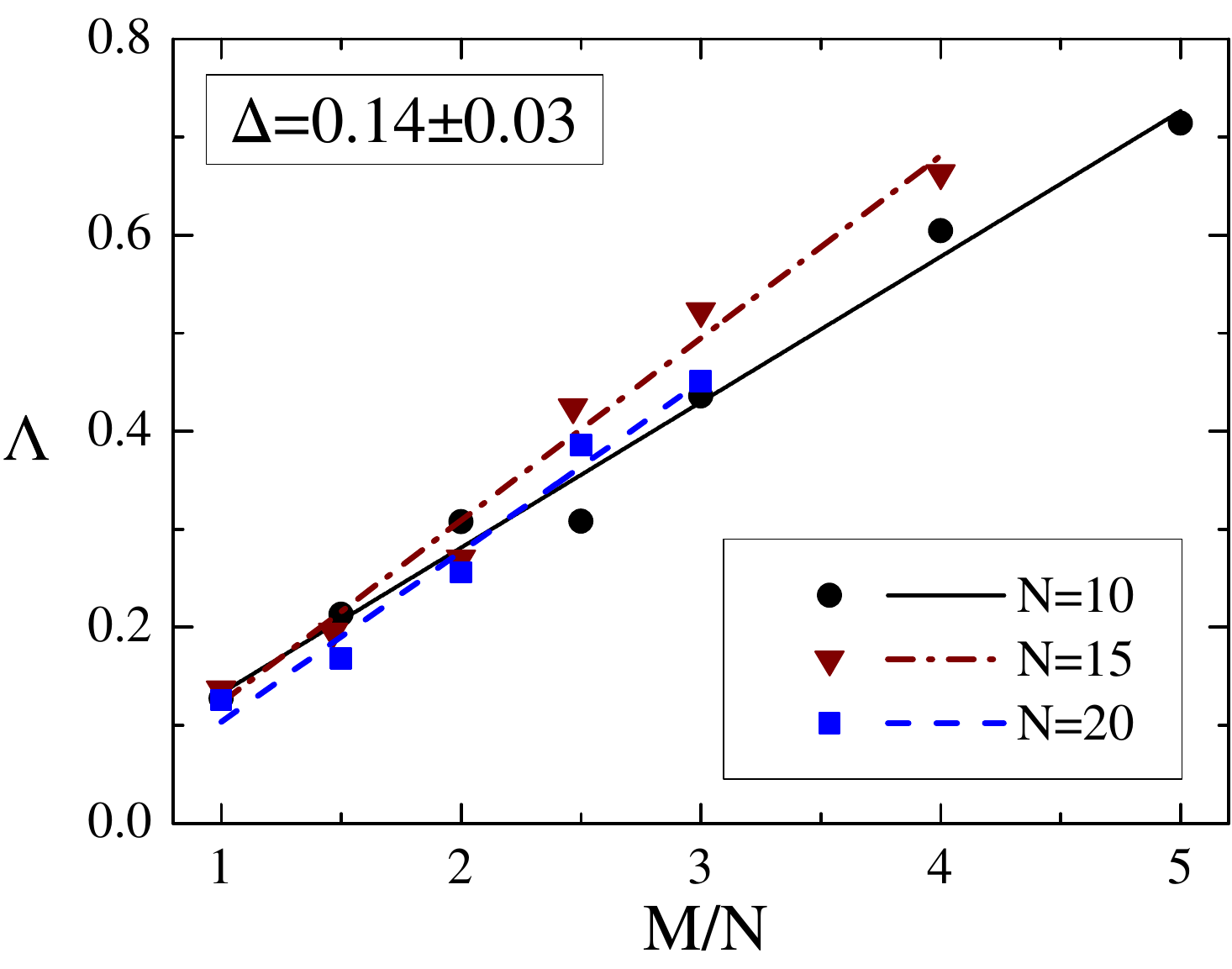}
  \caption{\label{h-Ising} Estimation of the minimal conformal weight in the Ising model.}
\end{figure}

\subsection{Conformal weights}
\label{sec:conformal-weights}

The second fitting procedure described above allows us to obtain the estimation for the minimal
conformal weight $h_{min}$ (or anomalous dimension $\Delta_{min}$). But it turns out that the obtained
result is very sensitive to the estimated value of $f(N)$. This is especially perceptible for large
torus size ratio $\frac MN\gg1$, when the difference between $f(N)$ and $f(N,M)$ is less than the
inaccuracy of the free energy value estimation. So one should exclude such lattices from a
consideration. On the other hand, non-minimal exponents become perceptible for the small ratio 
$\frac MN\approx1$. So we expect that a result of conformal weights estimation is far less accurate
than the usual precision of a critical exponents estimation.

With our precision, we obtain an acceptable result only for the Ising model. Fig. \ref{h-Ising}
shows the results for the quantity
\begin{equation}
  \label{eq:7}
  \Lambda\equiv-\frac{1}{2\pi}\ln\left(\frac{N^2(f(N)-f(N,M))}{T_c}\frac MN\right)\approx \frac{ M}{N} \Delta_{min},
\end{equation}
with a few values of lattice size $N$. $\Delta_{min}$ can be easily obtained by a linear fit. Then
we should average over different values of $N$.

If we wanted to obtain other conformal weights we would have to do multi-parametric fit with the
formula (\ref{eq:mainfit}). However such a fit would require very precise data that could be
obtained by increasing the number of algorithm iterations and enhancing the histogram flatness. This
makes simulations with the Wang-Landau algorithm very time-consuming. It may be more efficient to
use different approaches such as in the paper \cite{bastiaansen1998monte}.

\section{Results}
\label{sec:results}

\subsection{Ising model}
\label{sec:ising-model}

\begin{figure}[t]
  \centering
  \includegraphics[width=0.93\linewidth]{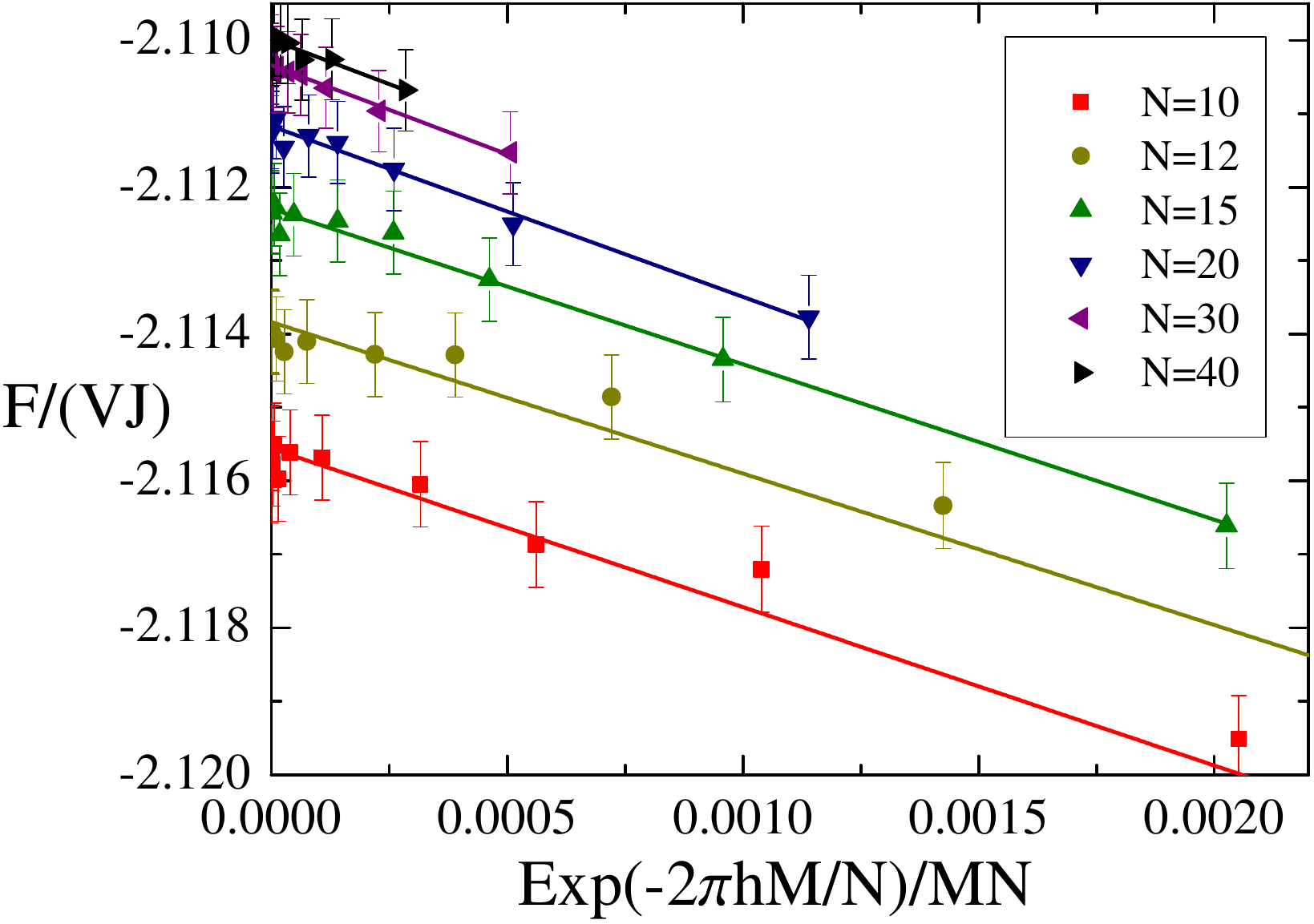}\\
  \includegraphics[width=0.95\linewidth]{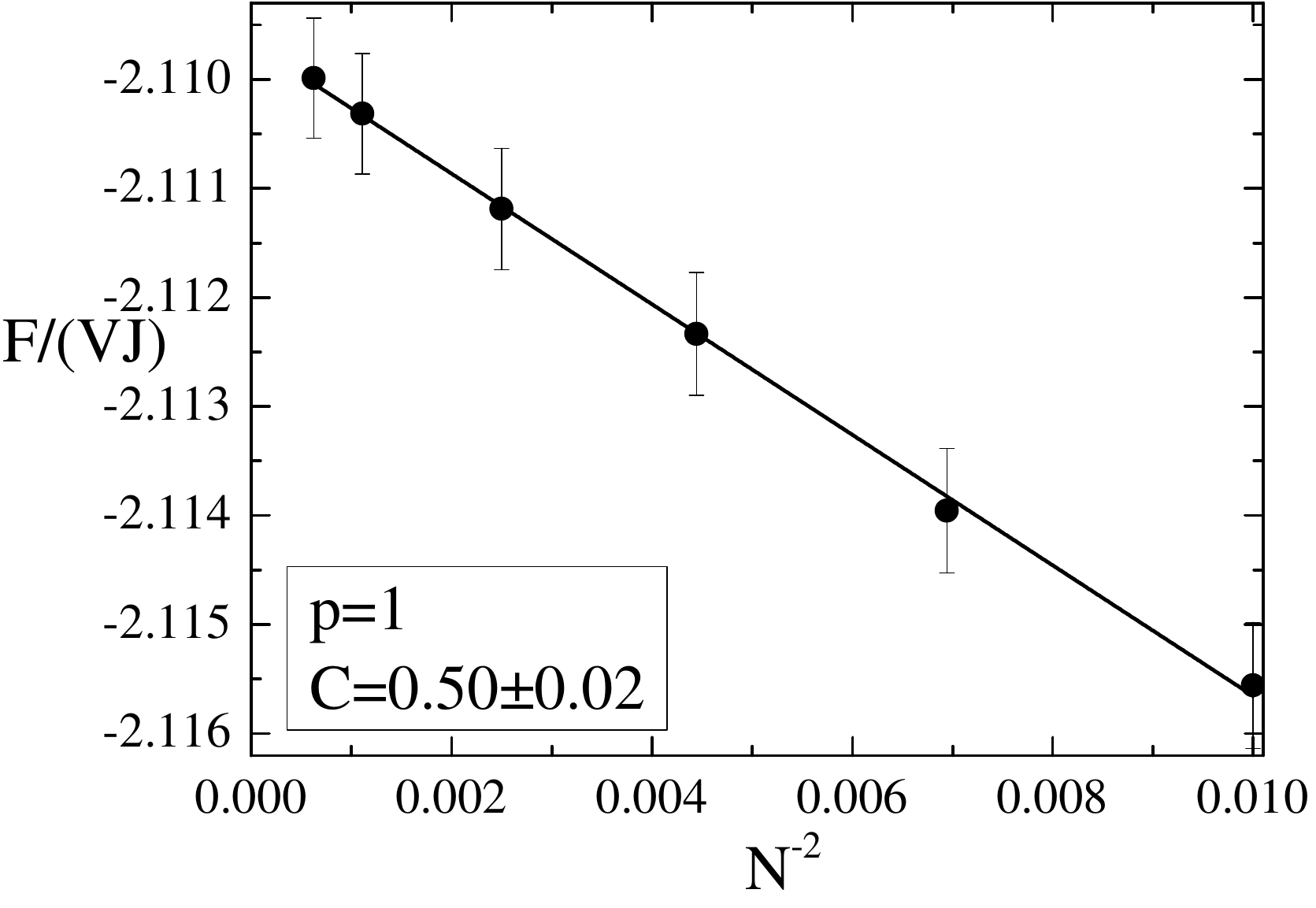}
  \caption{\label{fxnew} A finite-size scaling of the free energy $f(M,N)$ as a function of $q^{h}=e^{-2\pi h M/N}$ and $f(N)$ as a function of $N^{-2}$ in the Ising model.}
\end{figure}

The Hamiltonian of the Ising model \cite{lenz1920beitrag,ising1925beitrag} is: 
\begin{equation}
  H=-J\sum_{\left \langle i,j \right \rangle} s_{i}s_{j},\; s_{i}=\pm 1
\end{equation}
where $\left \langle i,j \right \rangle$ denotes the sum over neighbouring sites of a square lattice. The quantity $J$ is the exchange energy, and we set $J=1$ to fix the energy unit.

We use two values of the critical temperature that, in fact, give the same results for the
estimation of the central charge and the conformal weights. The first value is exact, obtained by Kramers-Wannier duality   \cite{kramers1941statistics}, whereas the
second one is obtained by Monte Carlo simulation with the cluster algorithm:
\begin{equation}
T_c\mathrm{(Exact)}=\frac{2}{\ln(\sqrt2+1)},\quad T_c\mathrm{(MC)}=2.2689(3).
\end{equation}

The critical point of the Ising model is described by the minimal model $\mathcal{M}(4,3)$ of the
conformal field theory \cite{belavin1984ics,difrancesco1997cft} with the central charge $c=\frac12$.
This model contains three primary fields with conformal dimensions $h_{r,s}$
\begin{equation}
h_{1,1}=0,\quad h_{2,1}=\frac1{16}, \quad h_{1,2}=\frac12.
\end{equation}
So, the minimal non-zero dimension is $\Delta_{min}=h_{min}+\bar h_{min}=\frac18$. Fitting by the
method discussed we obtain the value of $\Delta_{min}=0.14\pm 0.03$ (See Fig.\ref{h-Ising}), which
is in agreement with CFT. 

Our results on the finite-size scaling of the free energy are presented in
Fig.~\ref{fxnew}. From the extrapolation of $f(N,M)$ to $M\to\infty$ we have obtained the free
energy on the infinite cylinder and fitted the central charge. Our results are summarized in table
\ref{res-Ising}. They are in a good agreement with the well-known values of the central charge and
conformal weights obtained in CFT approach.
\begin{table}[h]
  \center
  \begin{tabular}{c|c|c}
     \hline
     \hline
        & Exact & This work\\
       \hline
       $c$ & $\frac12$ & $0.50\pm0.02$ \\
       $\Delta_{min}$ & $\frac18$ & $0.14\pm0.03$ \\
     \hline
     \hline
  \end{tabular}
  \caption{\label{res-Ising} Summary of the results for the Ising model.}
\end{table}

\subsection{Site-diluted Ising model}
\label{sec:site-diluted-ising}

Similar to the pure 2D Ising model, site-diluted Ising model is formulated on a lattice with
magnetic sites at the lattice vertices. We performed simulations on a triangular lattice, though the
critical behavior is independent of the lattice type. Each site has a spin $s=\pm 1$ or can be
non-magnetic ($s=0$). Sites are magnetic with the probability $p$, so the case of $p=1$ corresponds
to the Ising model. The Hamiltonian of the site-diluted model is the same as in the Ising model:
\begin{equation}
  H=-J\sum_{\left \langle i,j \right \rangle} s_{i}s_{j},\; s_{i}=\pm 1, \mbox{or}\;0
\end{equation}
where $\left \langle i,j \right \rangle$ denotes the sum over neighbouring sites of the lattice. 
The site-diluted Ising model has a critical point for any given value of $p$. The critical
temperature changes continuously with respect to $p$.

There is a claim that the central charge should also depend on $p$ \cite{najafi2016monte}. In our
study we found it not to be true. Our results and discussion are presented in the separate
publication \cite{Belov:2016lak}, here we present a brief summary.

We studied the model with the probabilities $p=0.8,\;0.9,\;0.95$ and performed
extensive simulations.

\begin{figure}[ht]
   \centering
   \includegraphics[width=0.95\linewidth]{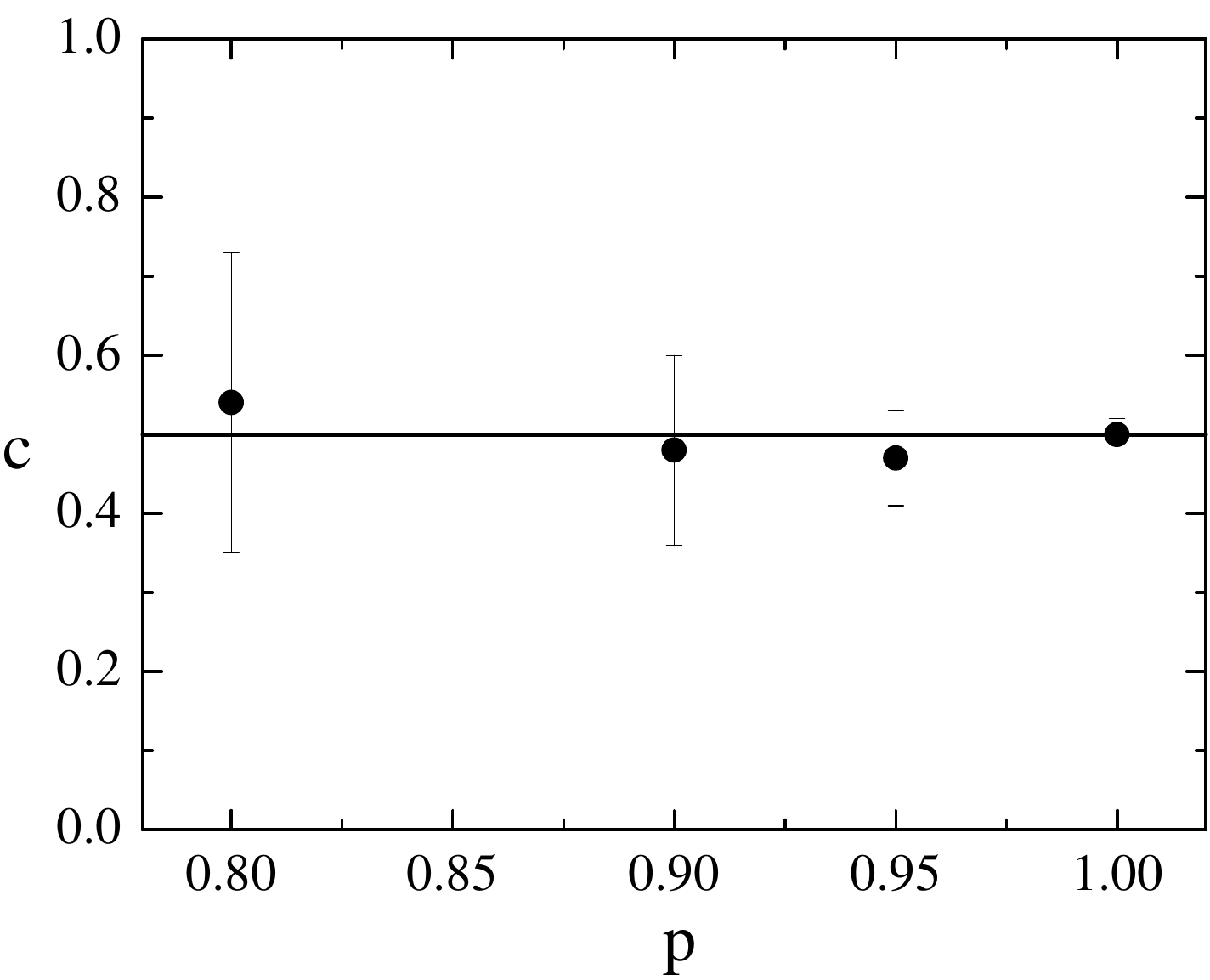}
   \caption{\label{cc09}Results for the central charge values in the site-diluted Ising model.}
\end{figure}

The results for the central charge are shown in Fig.~\ref{cc09} and in Table
\ref{results}. In contrast to Ref.~\cite{najafi2016monte}, we see that $c$ remains close to
$\frac{1}{2}$ as $p$ decreases, although the inaccuracy becomes larger.
\begin{table}[h]
  \center
  \begin{tabular}{c|c|c}
     \hline
     \hline
       $p$ & $T_c$ & $c$\\
       \hline
       $1$ & $3.64095\ldots$ & $0.50\pm0.02$ \\
       $0.95$ & $3.368\pm0.002$ & $0.47\pm0.06$ \\
       $0.9$ & $3.084\pm0.003$ & $0.48\pm0.12$ \\
       $0.8$ & $2.499\pm0.005$ & $0.54\pm0.19$ \\
     \hline
     \hline
  \end{tabular}
  \caption{\label{results} Summary of the results for the critical temperature and the central charge.}
\end{table}

This result is confirmed indirectly by results of the Wolff cluster
algorithm~\cite{wolff1989collective,landau2014guide}. The values of the critical indices for the
site-diluted Ising model agree with those for the pure Ising model as it is shown in
Ref.\cite{Belov:2016lak}. So, one can expect that for $0.65<p<1$ the site-diluted Ising model has
the Ising-like critical behavior with $c=\frac12$.

\subsection{Tricritical Ising model}
\label{sec:tricr-ising-model}

The tricritical Ising model is a short name for the tricritical point of the Blume-Capel model. The
Blume-Capel model was proposed to describe the behavior of Helium
\cite{blume1966theory,balbao1987operator,capel1966possibility}. It has a phase diagram with the
first and second order phase transition lines and the tricritical point. We study the model on the
square lattice in the tricritical point only.

The Hamilton of the model is similar to the Ising model:
\begin{equation}
  \label{eq:9}
  H=-J\sum_{<i,j>} s_{i} s_{j} +D \sum_{i} s_{i}^{2},
\end{equation}
where $<i,j>$ indicates nearest neighbours and spin $s_{i}$ takes values $-1,\;0,\;1$. This model is
different from the site-diluted Ising model. The distribution of the non-magnetic sites is not fixed
but is governed by the coupling constant $D$.

\begin{figure}[t]
  \centering
  \includegraphics[width=0.94\linewidth]{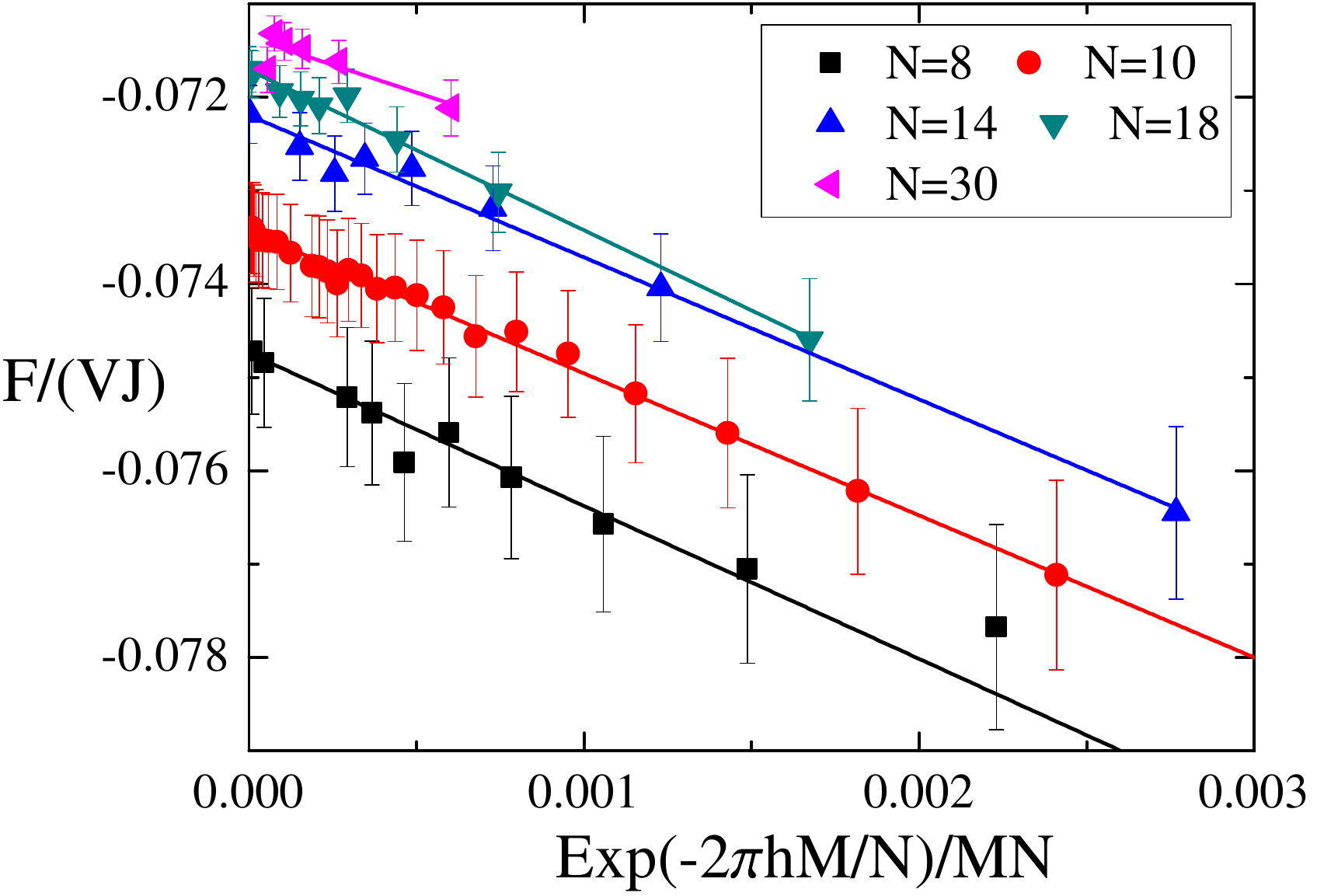}\\
  \includegraphics[width=0.95\linewidth]{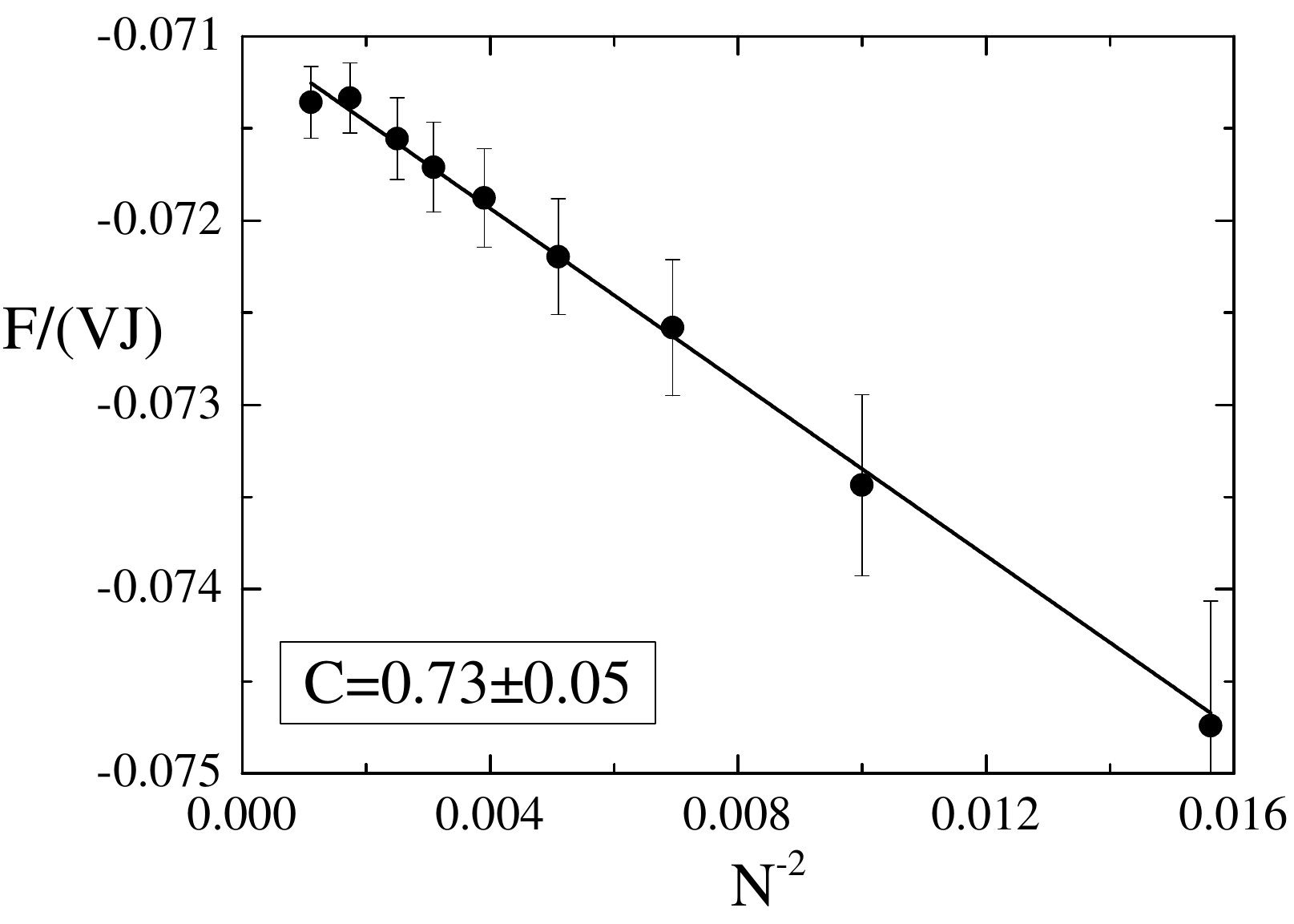}
  \caption{\label{ftricrit} A finite-size scaling of the free energy $f(M,N)$ as a function of $q^{h}=e^{-2\pi h M/N}$ and $f(N)$ as a function of $N^{-2}$ in the tricritical Ising model.}
\end{figure}
The tricritical point is at \cite{da2003global}
\begin{eqnarray}
\frac{D}{J}=1.9655,\quad T_c=0.610
\end{eqnarray}
It is described by the minimal model $\mathcal{M}(5,4)$ of the conformal field theory
\cite{belavin1984ics,difrancesco1997cft} with the central charge $c=\frac{7}{10}$. 

Our results are presented in fig. \ref{ftricrit}.
The estimation of the central charge by our method gives the value
\begin{equation}
c=0.73\pm0.05,
\end{equation}
that agrees with the exact result. 

\subsection{Three-state Potts model}
\label{sec:potts-model}

\begin{figure}[t]
  \centering
  \includegraphics[width=0.94\linewidth]{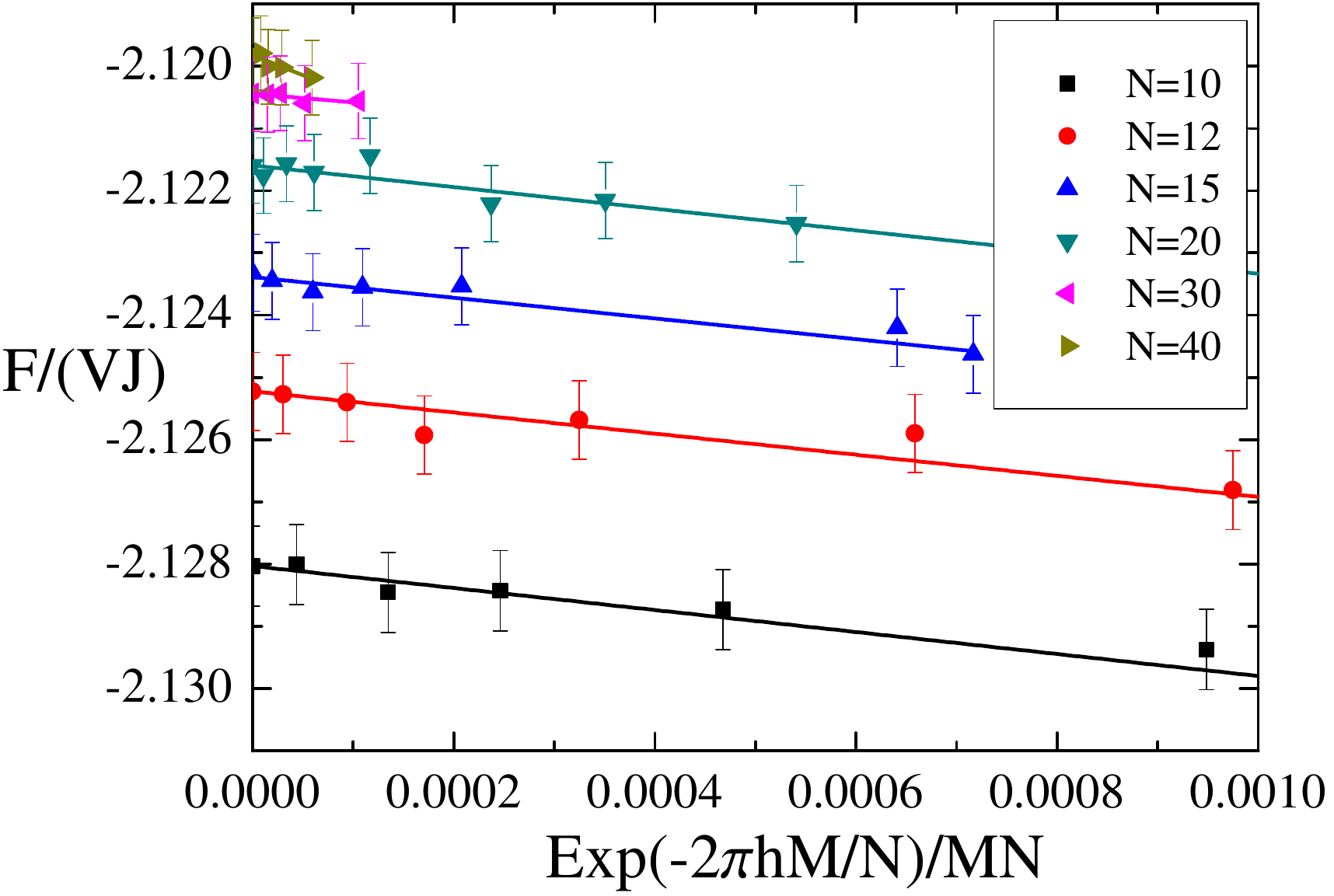}\\
  \includegraphics[width=0.95\linewidth]{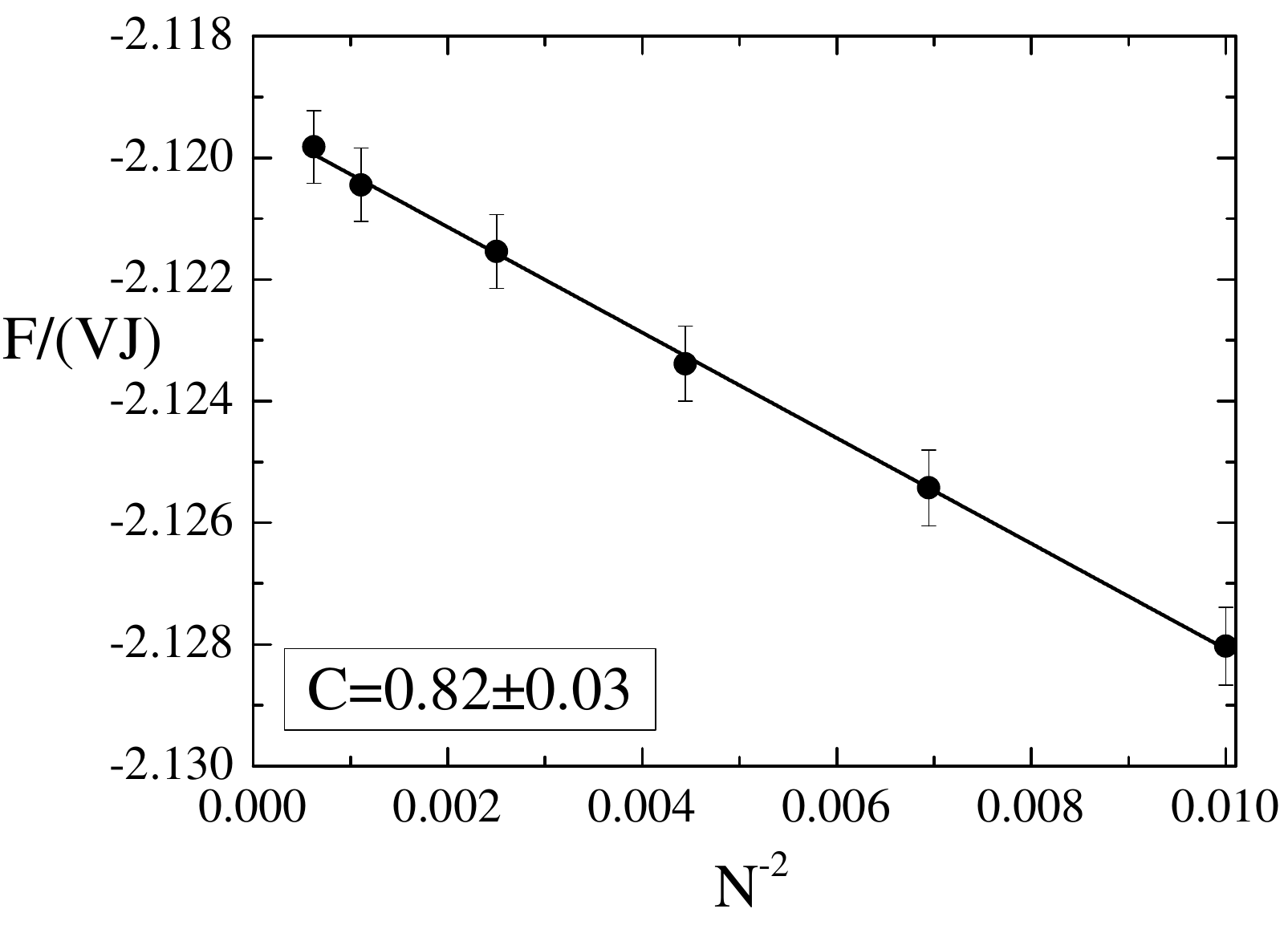}
  \caption{\label{fPotts3} A finite-size scaling of the free energy $f(M,N)$ as a function of $q^{h}=e^{-2\pi h M/N}$ and $f(\infty,N)$ as a function of $N^{-2}$ in the 3-state Potts model.}
\end{figure}

The Hamilton of the (non-planar) q-state Potts model \cite{potts1952some,PottsReview} is 
\begin{equation}
  \label{eq:Potts-ham}
  H=-J\sum_{<i,j>} \delta(s_{i}, s_{j}),\quad s=1,\,2,\ldots,q,
\end{equation}
where $\delta(s_{i},s_{j})$ is the Kronecker delta. 

As in the case of the Ising model ($q=2$), we consider two values of the critical temperature: the exact
\cite{potts1952some,PottsReview} ($q=3$) and the numerical one
\begin{equation}
T_c\mathrm{(Exact)}=\frac{2}{\ln(\sqrt q+1)},\quad T_c\mathrm{(MC)}=1.9897(3).
\end{equation}
The critical point of the three-state Potts model is described by the minimal model $\mathcal{M}(6,5)$ of the CFT 
\cite{dotsenko1984critical,difrancesco1997cft} with the central charge $c=\frac{4}{5}$.
Our simulations (see fig. \ref{fPotts3}) give the value for the central charge
\begin{equation}
c=0.82\pm0.03,
\end{equation}
that is also in agreement with the exact value.

\subsection{Four-state Potts model}
\label{sec:potts4-model}

Finally, we have also considered the 4-state Potts model. The critical point of it is not described by a
minimal model, but this model is a particular case of the Ashkin-Teller model \cite{ashkin1943statistics} with $c=1$. The
Hamiltonian is given by equation \eqref{eq:Potts-ham} with $q=4$. The numerical value of the critical
temperature is
\begin{equation}
T_c\mathrm{(MC)}=1.8204(3).
\end{equation}

\begin{figure}[t]
  \centering
  \includegraphics[width=0.94\linewidth]{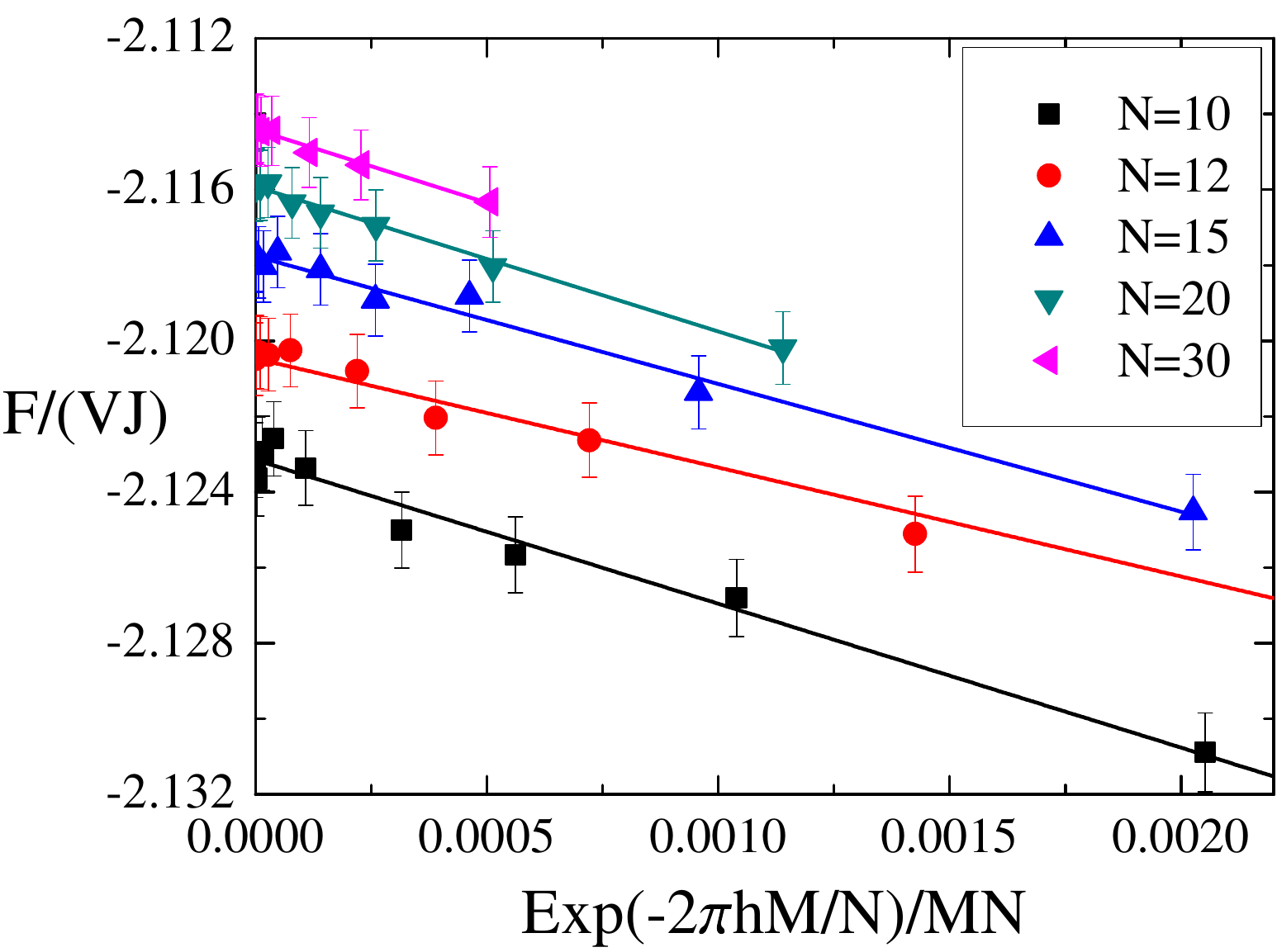}\\
  \includegraphics[width=0.95\linewidth]{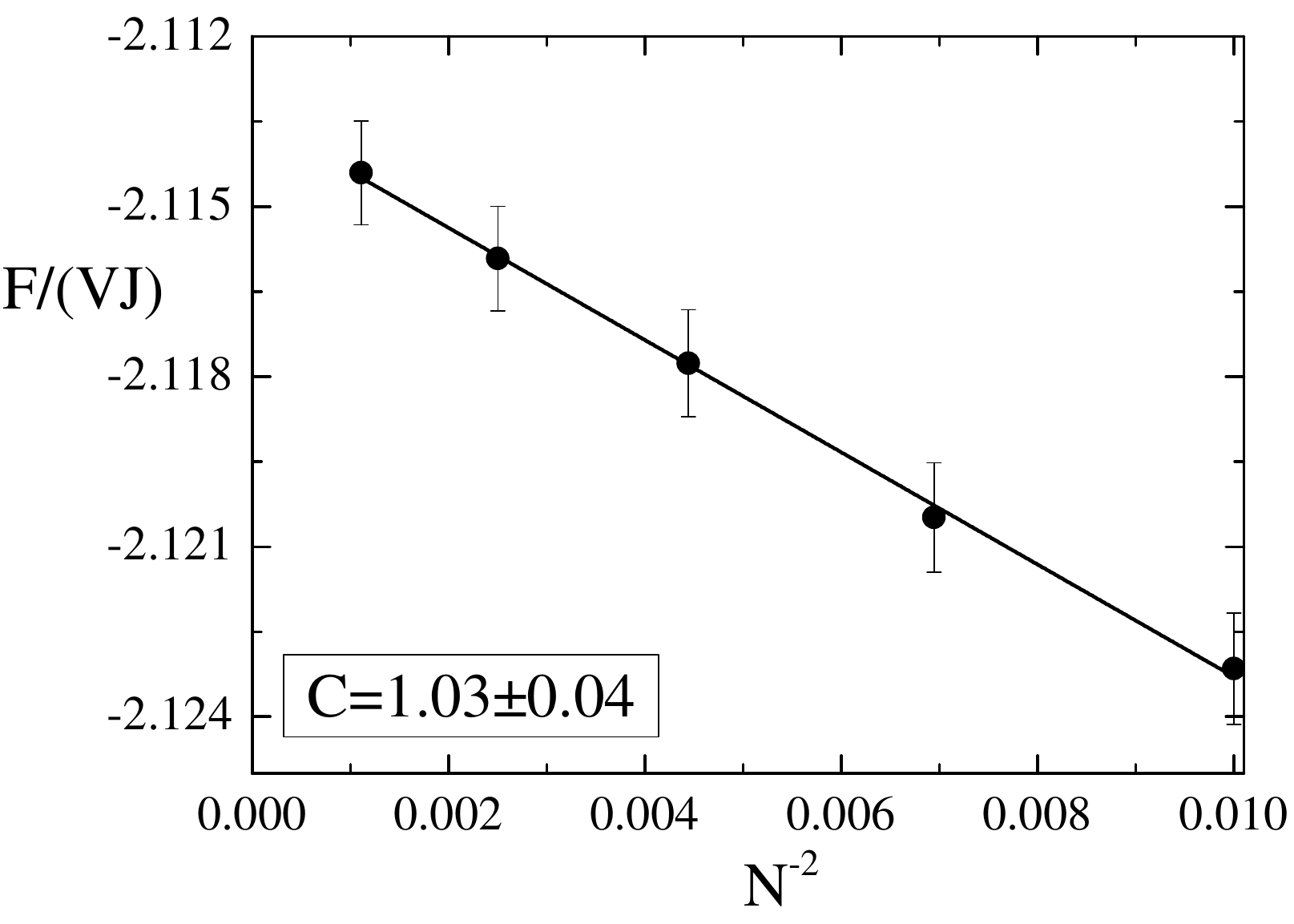}
  \caption{\label{fPotts4} A finite-size scaling of the free energy $f(M,N)$ as a function of $q^{h}=e^{-2\pi h M/N}$ and $f(N)$ as a function of $N^{-2}$ in the 4-state Potts model.}
\end{figure}

Using MC simulations, we obtained the following value of the central charge
\begin{equation}
c=1.03\pm0.04,
\end{equation}
as it is shown in Fig.~\ref{fPotts4}, that is close to 1.

\section*{Conclusion}
\label{sec:conclusion}

We have presented the simple method for estimation of the central charge in the CFT corresponding to
a two-dimensional lattice model at the critical point. The method is universal and can be
generalized also to non-discrete spin models. We have applied the method to the Ising, site-diluted
Ising, tricritical Ising, 3- and 4-state Potts models. Our numerical results on the central charge are
in a perfect agreement with the analytical results for the minimal models of the CFT.
We have also discussed a possibility of estimation of the conformal weights. It becomes possible if
one increases the number of algorithm iterations and enhances the histogram flatness. \medskip

\section*{Acknowledgments}
\label{sec:acknowledgements}

%
Anton Nazarov acknowledges the St. Petersburg State University for a support under the Research Grant No. 11.38.223.2015.
Alexander Sorokin is supported by the RFBR grant No. 16-32-60143.
The calculations were partially carried out using the facilities of the SPbU Resource Center ``Computational Center of SPbU''.

\bibliography{listing,bibliography}{} 
\bibliographystyle{apsrev4-1}

\end{document}